\documentclass[12pt]{article}
\usepackage{amsmath}
\usepackage{cite}
\begin{document}
\begin{flushright}
KANAZAWA-16-08\\
June, 2016
\end{flushright}
\vspace*{1cm}

\renewcommand\thefootnote{\fnsymbol{footnote}}
\begin{center} 
{\Large\bf Leptogenesis in a neutrino mass model coupled with inflaton}
\vspace*{1cm}

{\Large Daijiro Suematsu}\footnote[2]{e-mail:
~suematsu@hep.s.kanazawa-u.ac.jp}
\vspace*{0.5cm}\\

{\it Institute for Theoretical Physics, Kanazawa University, 
Kanazawa 920-1192, Japan}
\end{center}
\vspace*{1.5cm} 

\noindent
{\Large\bf Abstract}\\
We propose a scenario for the generation of baryon number asymmetry 
based on the inflaton decay in a radiative neutrino mass model 
extended with singlet scalars. 
In this scenario, lepton number asymmetry is produced through the decay 
of non-thermal right-handed neutrinos caused from the inflaton decay. 
Since the amount of non-thermal right-handed neutrinos could be much 
larger than the thermal ones, the scenario could work without any 
resonance effect for rather low reheating temperature. Sufficient 
baryon number asymmetry can be generated for much lighter right-handed 
neutrinos compared with the Davidson-Ibarra bound.  
             
\newpage
\setcounter{footnote}{0}
\renewcommand\thefootnote{\alph{footnote}}

\section{Introduction}
CMB observations suggest that there is an inflationary expansion era
in the early Universe \cite{uobs}. After inflation, the Universe should 
be thermalized enough to realize an initial stage of the hot Big-bang Universe.
Since inflation is usually assumed to be induced by the potential energy of 
a slow-rolling scalar field \cite{infrev}, this energy should be 
converted to radiation through so called reheating processes 
after the end of inflation. 
In order to make the reheating possible, inflaton should have some 
interactions with field contents of the standard model (SM) or others. 
As a result, in the effective model which is obtained after the inflaton 
is integrated out, its remnant is expected to be kept as 
effective interactions among the SM contents or additional fields 
at low energy regions. 
Since such interactions could be constrained by weak scale experiments 
or also be detected as some new physics at that scale,
their study is useful for the model building beyond the SM.
In particular, they might have some connection to the origin of the baryon 
number asymmetry in the Universe, which is one of big mysteries 
beyond the SM \cite{baryon}.     

In this paper, we study this issue assuming that inflaton is a singlet 
scalar of the SM gauge symmetry. In that case, only restricted couplings 
between the inflaton 
and ingredients of the model are allowed as renormalizable terms 
by the gauge symmetry. 
For example, we may consider a simple extension of the SM only with 
right-handed neutrinos $N_i$ and an additional doublet scalar.
In a supersymmetric case, it contains superpartners of the contents. 
In such a framework, a well-known example of the singlet inflaton is a 
sneutrino in the supersymmetric case. Sneutrino $\tilde N$ has a 
coupling $\tilde N\bar\ell\tilde \phi$, where $\ell$ and $\tilde\phi$ 
are the ordinary doublet lepton and a fermionic superpartner of 
the doublet Higgs scalar $\phi$, respectively.
In this model, reheating after the inflation and the associated 
generation of lepton number asymmetry due to this coupling has been 
studied in several articles \cite{sneut}.

In a non-supersymmetric case, gauge 
invariant renormalizable couplings of the singlet scalar $S$ 
with the contents of the model can be limited to two types if an additional
symmetry is imposed. These couplings are $S\bar N_iN^c_i$ 
and $S\eta^\dagger\phi$ where $\eta$ is the additional doublet scalar.
They are expected to bring about reheating and be relevant to
the generation of the baryon number asymmetry if $S$ plays a role of 
inflaton. A radiative neutrino mass model extended with singlet scalars is 
a typical example, which includes these couplings as 
phenomenologically important terms \cite{inf-nradm,bks,bks2}.
In this paper, we focus our study on such a model and 
propose a possible new scenario for the generation of the baryon 
number asymmetry through the reheating due to the above 
mentioned coupling.

\section{A radiative seesaw model extended by singlet scalars}
The radiative seesaw model \cite{ma} is a very simple but promising 
extension of the SM with an inert doublet scalar $\eta$ and right-handed 
singlet fermions $N_i$. 
They are assumed to have odd parity for an imposed $Z_2$ symmetry, 
although others are assigned even parity.
Lagrangian for these new fields contains the following terms,
\begin{eqnarray}
-{\cal L}&=&\sum_{i=1}^3\left[\sum_{\alpha=e,\mu,\tau}\Big(
-h_{\alpha i} \bar N_i\eta^\dagger\ell_\alpha
-h_{\alpha i}^\ast\bar\ell_\alpha\eta N_i \Big)+ 
\frac{1}{2}M_i \bar N_i^cN_i + \frac{1}{2}M_i\bar N_iN_i^c\right] \nonumber \\
&+&m_\phi^2\phi^\dagger\phi+m_\eta^2\eta^\dagger\eta 
+\lambda_1(\phi^\dagger\phi)^2 
+\lambda_2(\eta^\dagger\eta)^2 
+\lambda_3(\phi^\dagger\phi)(\eta^\dagger\eta)  
+\lambda_4(\eta^\dagger\phi)(\phi^\dagger\eta)  \nonumber \\ 
&+&\frac{\lambda_5}{2}\left[(\eta^\dagger\phi)^2 +(\phi^\dagger\eta)^2 \right], 
\label{model}
\end{eqnarray}
where $\ell_\alpha$ is a left-handed doublet lepton and $\phi$ is 
the ordinary doublet Higgs scalar. The coupling constants 
$\lambda_i$'s are real. 
The model is known to give a simultaneous explanation for the existence of 
neutrino masses and dark matter (DM) \cite{f-nradm,tribi}. 
Neutrino masses are induced at one-loop level and 
DM is prepared as the lightest $Z_2$ odd field.
Moreover, the model can also explain the baryon number asymmetry 
in the Universe through leptogenesis if the masses of $N_i$ are finely 
degenerate \cite{ks}.

First, we briefly overview these features.
For the definite argument, we assume that the lightest $Z_2$
odd field is a lightest neutral component of $\eta$ here. 
Its mass is expressed as 
$M_\eta^2=m_\eta^2+(\lambda_3+\lambda_4+\lambda_5)\langle\phi\rangle^2$ and
it is taken to be of $O(1)$ TeV.
Since the SM contents are assigned even parity, it is stable and then it can 
be a good DM candidate \cite{inert,ham}. 
In fact, it is known to realize the required DM relic abundance 
only by fixing the couplings $\lambda_{3,4}$ at suitable values \cite{ks}. 
The neutrino oscillation data could also be roughly 
explained by assuming a simple flavor structure such as\cite{tribi}
\begin{equation}
h_{ei}=0, \quad h_{\mu j}=h_{\tau j}\equiv h_j  \quad (j=1,2); \qquad
h_{e3}=h_{\mu 3}=-h_{\tau 3}\equiv h_3.
\label{flavor}
\end{equation}
In this case, the neutrino mass matrix can be written as
\begin{equation}
{\cal M}=(h_1^2\Lambda_1+ h_2^2\Lambda_2)
\left(\begin{array}{ccc}0&0&0\\0&1&1\\0&1&1\\\end{array}\right)+
h_3^2\Lambda_3\left(\begin{array}{ccc}1&1&-1\\1&1&-1\\-1&-1&1\\
\end{array}\right),
\end{equation} 
and $\Lambda_i~(i=1-3)$ is given by
\begin{equation}
\Lambda_i=\frac{\lambda_5\langle\phi\rangle^2}{8\pi^2}
\frac{M_i}{M_\eta^2-M_i^2}
\left(1+\frac{M_i^2}{M_\eta^2-M_i^2}\ln\frac{M_i^2}{M_\eta^2}\right).
\end{equation} 
The mass eigenvalues of this matrix are obtained as
\begin{eqnarray}
&&m_1=0,  \qquad m_2= 3|h_3|^2\Lambda_3,  \nonumber \\
&&m_3= 2\Big[|h_1|^4\Lambda_1^2+|h_2|^4\Lambda_2^2+
2|h_1|^2|h_2|^2\Lambda_1\Lambda_2\cos 2(\theta_1-\theta_2)\Big]^{1/2}, 
\end{eqnarray}
where $\theta_j= {\rm arg}(h_j)$.
If we apply the parameters shown in Table 1 to this mass matrix as examples, 
the neutrino oscillation parameters required for the normal hierarchy 
can be obtained except that the tri-bimaximal PMNS matrix is brought 
about \cite{tribi}.\footnote{Although we can fix the parameters 
at the values required for the inverted hierarchy in the similar way, 
we confine the present study to the normal hierarchy.}
Since we now know that $\theta_{13}$ has a non-zero value, 
we have to modify the flavor structure given in eq.~(\ref{flavor}) \cite{ks}. 
However, since the required modification is expected to cause no 
crucial effect to the leptogenesis scenario, the use of this simple 
flavor structure is enough for the present purpose. 
Although the resonant leptogenesis works in this model, unnatural 
fine degeneracy among the right-handed neutrino masses seems to be 
required \cite{ks}.\footnote{This mass degeneracy might be explained 
by assuming the pseudo-Dirac nature for $N_i$, which could be caused by
symmetry breaking at a TeV scale \cite{ks1}.} 
We consider an extension of the model with singlet scalars, 
which could remedy this fault without spoiling the favorable features 
of the model mentioned above. 

\begin{figure}[t]
\begin{center}
\begin{tabular}{c|cccccccc}\hline
& $\lambda_5$ &$|h_1|$ & $|h_2|$ & $|h_3|$ & $M_1$ & $M_2$ & $M_3$ & $M_\eta$ 
\\\hline  
(a)&$4\times 10^{-4}$& $10^{-4}$ & $5.67\times 10^{-3}$ & $2.63\times 10^{-3}$ & 
$10^6$ & $3\times 10^6$ & $6\times 10^6$ & $10^3$ \\ 
(b)&$3\times 10^{-5}$ &$10^{-4}$ & $3.33\times 10^{-3}$& $1.45\times 10^{-3}$ & 
$10^4$& $3\times 10^4$& $6\times 10^4$& $10^3$ \\
\hline 

\end{tabular}
\end{center} 
\vspace*{3mm}

{\footnotesize Table 1~ Typical parameter sets for the neutrino mass 
generation, which explain the neutrino oscillation data. 
A GeV unit is used for the mass.} 
\end{figure}

In the model defined by eq.~(\ref{model}), 
we can suppose two types of lepton number assignment for the new fields 
such as
(i)~$L(N_i)=0$ and $L(\eta)=1$, in which the lepton number is violated 
through the $\lambda_5$ terms, and
(ii)~$L(N_i)=1$ and $L(\eta)=0$, in which the lepton number is violated 
through the mass terms of $N_i$.
If these lepton number violating terms are supposed to have its origin 
at high energy regions and they are effectively induced from it 
at low energy regions, some new fields might be introduced to give their origin.
Along this idea, we consider an extension of the model with singlet 
scalars $S_a$. 
The addition of $S_a$ allows to introduce gauge invariant terms such as 
$\mu_a S_a\eta^\dagger\phi$ and $y_i^{(a)}S_a\bar N_iN_i^c$. 
As is shown below, these could induce the above mentioned lepton 
number violating terms as the effective ones in the case (i) and (ii), 
respectively.
Since we impose the lepton number conservation in these terms,
$S_a$ should be assigned the lepton number +1 and $-2$ in each case.
In order to keep important features of the original model, 
the $Z_2$ symmetry should be an exact symmetry.  
Thus, in the former case, $S_a$ should have odd $Z_2$ parity and 
$\langle S_a\rangle=0$.
If $S_a$ is supposed to be heavy enough and it is integrated out 
to derive the low energy effective model, the $\lambda_5$ term 
in eq.~(\ref{model}) is induced as long as a lepton number 
violating term $m_a^2S_a^2$ exists \cite{inf-nradm}. 
On the other hand, in the latter case, $S_a$ should have even parity of 
$Z_2$ and then $\langle S_a\rangle\not=0$ can be allowed at TeV or 
higher energy scales. In such a case, even if $M_i=0$ is supposed 
in eq.~(\ref{model}), this vacuum expectation value generates 
Majorana masses for $N_i$ without violating the $Z_2$ symmetry. 

We identify the one of these singlet scalars $S_a$ with inflaton.
It is discussed that it could play a role of inflaton by assuming 
special potential for it or its non-minimal coupling with Ricci 
scalar \cite{inf-nradm,bks,bks2}.
In the model corresponding to the case (i), both the inflation 
and the non-thermal leptogenesis associated to the reheating 
due to the inflaton decay through the interaction 
$\mu_a S_a\phi^\dagger\eta$ has already been discussed in \cite{bks2}. 
The model is found to work well under the appropriate conditions there.   
In this paper, we study another possibility in the case (ii), where 
the reheating is caused by the coupling $S_a\bar N_iN_i^c$. 
Leptogenesis is also supposed to be brought about through this reheating 
process.
In this direction, different types of scenario for the leptogenesis 
might be considered depending on the way how the lepton number asymmetry 
is generated in the doublet lepton sector.
Here, we consider that the lepton number asymmetry 
is produced through the decay of $N_i$ which is non-thermally 
produced through the inflaton decay.\footnote{Another scenario might 
be constructed by assuming that the lepton number asymmetry is 
produced directly through the inflaton decay to the Dirac type 
right-handed neutrinos $N_i$ under the condition $M_i=0$. 
Such a possibility will be discussed elsewhere.}

The model might contain terms relevant to the singlet scalars $S_a$ as 
\begin{eqnarray}
-{\cal L}&=& \tilde m_{S_a}^2S_a^\dagger S_a+\kappa_{S}^{(a)}(S_a^\dagger S_a)^2 
+\kappa_\phi^{(a)}(S_a^\dagger S_a)(\phi^\dagger\phi)
+\kappa_\eta^{(a)}(S_a^\dagger S_a)(\eta^\dagger\eta) \nonumber \\
&+&y_i^{(a)}S_a\bar N_i^cN_i 
+ y_i^{(a)\ast}S_a^\dagger \bar N_iN_i^c \nonumber \\
&+&\frac{1}{2}m_{S_a}^2S_a^2 +\frac{1}{2}m_{S_a}^2S_a^{\dagger 2}. 
\label{smodel}
\end{eqnarray}
The lepton number is explicitly broken through both the Majorana masses of 
$N_i$ in eq.~(\ref{model}) and mass terms of $S_a$ in the third line of 
eq.~(\ref{smodel}). The latter one makes the components of $S_a$ 
split into mass eigenstates $S_{\pm a}$ with mass eigenvalues  
$m_{\pm a}^2=m_{S_a}^2\pm\tilde m_{S_a}^2$. As is easily found, $S_{+a}$ and 
$S_{-a}$ correspond to real and imaginary parts of $S_a$, respectively.  
This results in the lepton number violation in the Yukawa coupling 
$\frac{y_i^{(a)}}{\sqrt 2}S_{\pm a}\bar N_iN_i^c$.
We do not consider the spontaneous mass generation for $N_i$ 
through the interaction given in the second line and then 
$\langle S_a\rangle=0$ is supposed here. 
This extension could change phenomenology in the original Ma model.
The $\kappa_\phi^{(a)}$ and $\kappa_\eta^{(a)}$ terms could affect 
the quartic couplings $\lambda_1$-$\lambda_4$ for $\phi$ and $\eta$ 
through the radiative effects. 
As a result, they might be constrained by weak scale experiments. 
On the other hand, $\frac{y_i^{(a)}}{\sqrt 2}S_{\pm a}\bar N_iN_i^c$ could 
be relevant to the leptogenesis. 
In the following parts, we focus our discussion on this latter point.

\section{Non-thermal leptogenesis associated to reheating}
We assume that a real component of $S_1$ plays a role of inflaton. 
It is represented as $S_{\rm inf}(\equiv S_{+1})$ in the following part.
When the inflation ends, $S_{\rm inf}$ is supposed to start damping 
oscillation around a potential minimum $\langle S_{\rm inf}\rangle=0$.  
At the first stage of this oscillation, its amplitude is large and then 
preheating could occur through the quartic couplings 
$\frac{\kappa_\phi}{2} S_{\rm inf}^2\phi^\dagger\phi$ and 
$\frac{\kappa_\eta}{2} S_{\rm inf}^2\eta^\dagger\eta$ \cite{pre,reheat}.
Although $\phi$ and $\eta$ might be produced explosively through 
the resonance effect for suitable values of $\kappa_\phi$ and $\kappa_\eta$, 
the following decay of $\phi$ and $\eta$ 
cannot produce any lepton and baryon number 
asymmetry. This situation is not changed 
even if $\eta$ is heavier than $N_i$.
Although $\eta$ can decay into $\ell_\alpha N_i$, any lepton number asymmetry 
is not generated in the doublet lepton sector through this decay because 
of the cancellation of the asymmetry between the yields 
from $\eta$ and $\eta^\dagger$.

At the later stage of this oscillation, 
the inflaton decay is expected to be induced by the coupling 
$\frac{y_i}{\sqrt 2} S_{\rm inf}\bar N_iN_i^c$ where the coupling 
$y_i^{(1)}$ is abbreviated to $y_i$. 
The inflaton energy is expected to be converted dominantly 
to one of the right-handed neutrinos $N_i$, which has the mass 
satisfying $M_i<\frac{m_{S_{\rm inf}}}{2}$ and also the largest 
partial decay width
\begin{equation}
\Gamma_{S_{\rm inf}}^{(i)}=\frac{|y_i|^2}{8\pi}m_{S_{\rm inf}}
\left(1-\frac{4M_i^2}{m_{S_{\rm inf}}^2}\right)^{1/2}.
\label{swidth}
\end{equation}
If we fix such a $N_i$ at $N_1$ for the concreteness\footnote{In this study, 
$N_1$ is assumed to be the lightest one as shown in Table~1, which is favored 
to suppress the washout of the generated lepton number asymmetry as
discussed later.},
the resulting number density of $N_1$ produced through this 
decay can be estimated as 
\begin{equation}
n_{N_1}^{\rm nonth}=\frac{\rho_{S_{\rm inf}}}{M_1}
=\frac{3|y_1|^4}{64\pi^2}\frac{M_{\rm pl}^2m_{S_{\rm inf}}^2}{M_1}
\left(1-\frac{4M_1^2}{m_{S_{\rm inf}}^2}\right),
\label{num-N}
\end{equation} 
where we use a value of the inflaton energy density 
$\rho_{S_{\rm inf}}$. It is fixed through the condition 
$H\simeq \Gamma_{S_{\rm inf}}^{(1)}$ for the Hubble parameter 
$H^2=\frac{\rho_{S_{\rm inf}}}{3M_{\rm pl}^2}$.

If the decay rate of $N_1$ is larger than the one of $S_{\rm inf}$, 
the produced $N_1$ is expected to decay to $\ell_\alpha\eta$ immediately
since it is the lowest order process. 
In such a case, the decay of $N_1$ is considered to occur in a non-thermal 
situation before the completion of thermalization.
Since the decay width of $N_1$ is estimated under the assumption 
(\ref{flavor}) for the neutrino Yukawa couplings as 
\begin{equation}
\Gamma_{N_1}=
\frac{|h_1|^2}{4\pi}M_1
\left(1-\frac{M_\eta^2}{M_1^2}\right),
\label{width}
\end{equation}
the required condition $\Gamma_{S_{\rm inf}}^{(1)} < \Gamma_{N_1}$ might 
be roughly expressed as
\begin{equation}
\left(\frac{|y_1|}{10^{-8}}\right)
\left(\frac{m_{S_{\rm inf}}}{10^7~{\rm GeV}}\right)^{\frac{1}{2}} < 
10^3\left(\frac{|h_1|}{10^{-3}}\right)
\left(\frac{M_1}{10^3~{\rm GeV}}\right)^{\frac{1}{2}}
\left(1-\frac{M_\eta^2}{M_1^2}\right)^{\frac{1}{2}}
\left(1-\frac{4M_1^2}{m_{S_{\rm inf}}^2}\right)^{-\frac{1}{4}}.
\label{inst}
\end{equation}  
Here we note that the inflaton mass $m_{S_{\rm inf}}$ is not constrained by
the observational data of CMB as long as we assume the suitable 
inflation scenario such as the ones discussed in \cite{bks,bks2}.

The reheating temperature $T_R$ is estimated from 
$H\simeq\Gamma_{S_{\rm inf}}^{(i)}$ as\footnote{The reheating temperature should be 
estimated by using the $S_{\rm inf}$ decay rate to the final states composed of 
four particles $2(\bar\ell_\alpha\eta)$ instead of eq.~(\ref{swidth}). 
However, both of them give the same value for $T_R$ 
as long as the condition (\ref{inst}) is imposed. 
As a result, no $h_1$ dependence appears in the expression of $T_R$.}
\begin{equation}
T_R\simeq 5.3\times 10^3\left(\frac{|y_1|}{10^{-8}}\right)
\left(\frac{m_{S_{\rm inf}}}{10^7~{\rm GeV}}\right)^{\frac{1}{2}}
\left(1-\frac{4M_1^2}{m_{S_{\rm inf}}^2}\right)^{\frac{1}{4}}~{\rm GeV}.
\label{tr}
\end{equation}
The reheating temperature is found to take a fixed value for a constant 
value of $|y_1|^2m_{S_{\rm inf}}$.
Since we suppose that the DM abundance is realized by the 
thermal relic of the lightest neutral component of $\eta$ in this model, 
$T_R>M_\eta$ should be fulfilled.
This requires 
\begin{equation}
\left(\frac{|y_1|}{10^{-8}}\right)
\left(\frac{m_{S_{\rm inf}}}{10^7~{\rm GeV}}\right)^{\frac{1}{2}}
>0.2\left(\frac{M_\eta}{10^3~{\rm GeV}}\right).
\end{equation}
An interesting thing is that the number density (\ref{num-N}) 
could be much larger than 
the thermal equilibrium value even for the relativistic $N_1$ as
\begin{equation}
\frac{n_{N_1}^{\rm nonth}} {n_{N_1}^{\rm th}}\simeq 10^3 
\left(\frac{|y_1|}{10^{-8}}\right)
\left(\frac{m_{S_{\rm inf}}}{10^7~{\rm GeV}}\right)^{\frac{1}{2}}
\left(\frac{10^3~{\rm GeV}}{M_1}\right)
\left(1-\frac{4M_1^2}{m_{S_{\rm inf}}^2}\right)^{\frac{1}{4}},
\label{enhance}
\end{equation}
where we use eqs.~(\ref{num-N}) and (\ref{tr}). 
If $N_1$ is non-relativistic and $T_R$ is much smaller than $M_1$,
this ratio is enhanced by a factor $e^{\frac{M_1}{T_R}}$.

\input epsf
\begin{figure}[t]
\begin{center}
\epsfxsize=14cm
\leavevmode
\epsfbox{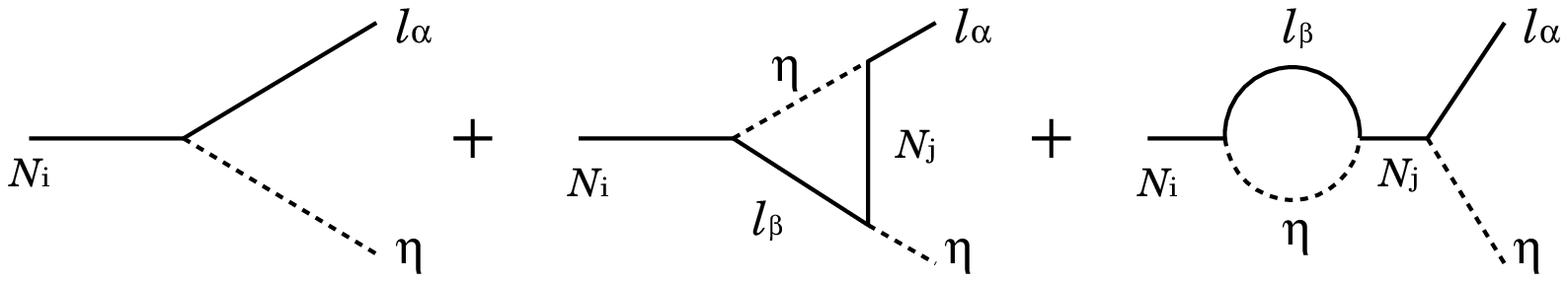}

{\footnotesize {\bf Fig.~1}~~Feynman diagrams contributing to 
the generation of the lepton number asymmetry. }
\end{center}
\end{figure}

Since the $N_1$ decay to $\ell_\alpha\eta$ can satisfy the Sakharov conditions, 
the lepton number asymmetry could be generated through 
this process. In could be estimated as\footnote{This $Y_L$ should be 
understood as $Y_{B-L}$ under the existence of sphaleron interaction.}
\begin{equation}
Y_L= 2 \varepsilon \frac{n_{N_1}}{s},
\label{ini-yl}
\end{equation}
where $Y_L$ is defined as 
$Y_L=\sum_\alpha \frac{n_{\ell_\alpha} -n_{\bar\ell_\alpha}}{s}$ by using
the entropy density $s$.  The $CP$ 
asymmetry in the decay $N_1\rightarrow \sum_{\alpha} \ell_\alpha\eta$
is represented by $\varepsilon$.
It is brought about by the interference between tree and 
one-loop diagrams shown in Fig.~1 and can be derived as \cite{cp}
\begin{eqnarray}
\varepsilon&=&\frac{\Gamma(N_1\rightarrow\sum_\alpha\ell_\alpha\eta^\dagger)-
\Gamma(N_1\rightarrow\sum_\alpha\bar\ell_\alpha\eta)}
{\Gamma(N_1\rightarrow\sum_\alpha\ell_\alpha\eta^\dagger)
+\Gamma(N_1\rightarrow\sum_\alpha\bar\ell_\alpha\eta)}  \nonumber \\
&=&\frac{1}{16\pi C}\frac{\sum_{j=2,3}{\rm Im}
\left[\left(\sum_{\alpha=e,\mu,\tau}h_{\alpha 1}h^\ast_{\alpha j}\right)^2\right]}
 {\sum_{\alpha =e,\mu,\tau}h_{\alpha 1}h_{\alpha 1}^\ast}
~G\left(\frac{M_j^2}{M_1^2},\frac{M_\eta^2}{M_1^2}\right), 
\label{cp}
\end{eqnarray}
where $C=\frac{3}{4}+\frac{1}{4}\left(1-\frac{M_\eta^2}{M_1^2}\right)^2$ 
and $G(x,y)$ is defined by
\begin{eqnarray} 
G(x,y)&=&\frac{5}{4}F(x,0)+\frac{1}{4}F(x,y)
+\frac{1}{4}(1-y)^2\left[F(x,0)+F(x,y)\right], \nonumber \\
F(x,y)&=&\sqrt{x}\left[1-y-(1+x)\ln\left(\frac{1-y+x}{x}\right)\right].
\end{eqnarray} 
If we apply the flavor structure of neutrino Yukawa couplings 
given in eq.~(\ref{flavor}) to this formula, 
$\varepsilon$ is expressed as
\begin{equation}
\varepsilon=\frac{|h_2|^2\sin 2(\theta_1-\theta_2)}{8\pi C}
~G\left(\frac{M_2^2}{M_1^2},\frac{M_\eta^2}{M_1^2}\right).
\end{equation}  
We assume the maximum $CP$ phase 
$\sin 2(\theta_1-\theta_2)=1$ in the following numerical study.

When the reheating completes through the inflaton decay, 
all fields could be considered to take the thermal distribution at 
the temperature $T_R$. However, the asymmetry produced through
the decay of the non-thermal $N_1$ could exist as 
$Y_L=2\varepsilon\frac{n^{\rm nonth}_{N_1}}{s}$ at this stage. 
If we take this view point, this asymmetry could be treated as its 
initial value at the reheating temperature $T_R$ for the following 
evolution of $Y_L$.
In the usual thermal leptogenesis scenario discussed 
in \cite{ks}, $N_1$ is considered to be in the thermal equilibrium
due to the assumption $T_R> M_1$.
Thus, we find the relation at $T_R$, by comparing these two cases, such as
\begin{equation}
Y_L^{\rm nonth}=\frac{n^{\rm nonth}_{N_1}}{n^{\rm th}_{N_1}}Y_L^{\rm th}.
\end{equation} 
This suggests that $Y_L^{\rm nonth}$ could have a largely enhanced value 
compared with $Y_L^{\rm th}$ as long as the factor
$\frac{n^{\rm nonth}_{N_1}}{n^{\rm th}_{N_1}}$ takes an enhanced value as suggested 
in eq.~(\ref{enhance}).
However, we should note that this enhanced initial asymmetry can play a
substantial role for the generation of the sufficient baryon number asymmetry 
only if the washout of the lepton number asymmetry is ineffective 
at a neighborhood of $T_R$. Such a situation could be realized 
owing to the Boltzmann suppression only for $M_1>T_R$. 
By combing it with the requirement 
$\frac{n^{\rm nonth}_{N_1}}{n^{\rm th}_{N_1}}\gg 1$,
the condition might be expressed as
\begin{equation}
10^{-3}\left(\frac{M_1}{10^3~{\rm GeV}}\right)
\left(1-\frac{4M_1^2}{m_{S_{\rm inf}}^2}\right)^{-\frac{1}{4}}
\ll\left(\frac{|y_1|}{10^{-8}}\right)
\left(\frac{m_{S_{\rm inf}}}{10^7~{\rm GeV}}\right)^{\frac{1}{2}}
<0.2 \left(\frac{M_1}{10^3~{\rm GeV}}\right).
\label{supp}
\end{equation}
We note that it is easy for this condition to be consistent with
eq.~(\ref{inst}) as long as the Yukawa coupling $h_1$ is larger than 
$10^{-6}$.   
Thus, we can expect that the lepton number asymmetry, which is enhanced 
from that in the thermal leptogenesis, could be obtained
for the model parameters which are suitably fixed without any serious tuning.

A typical situation is that the non-relativistic $N_1$ is 
produced in such a circumstance that lepton number violating processes 
freeze out. In that case, the initial lepton asymmetry could be kept until
the weak scale. On the other hand, if the washout effects are in thermal 
equilibrium, the initial lepton asymmetry is immediately erased and 
the scenario reduces to the case similar to the usual thermal leptogenesis.
This feature makes a rather low reheating temperature favorable 
in this scenario.
If this favorable situation is realized, the sufficient lepton number 
asymmetry is expected to be generated from the right-handed neutrino
whose mass is much smaller than the Davidson-Ibarra bound \cite{di}
without any resonance effect \cite{resonant}. 
It might give an another interesting possibility for the 
leptogenesis in the radiative neutrino mass model. 

For a quantitative check of the above discussion, the analysis of the 
Boltzmann equations is required to estimate the washout effect of the 
generated lepton number asymmetry, especially, in a marginal situation.
The lepton number asymmetry given in eq.~(\ref{ini-yl}) could be 
affected by the washout through lepton number violating scattering 
and the inverse decay at a neighborhood of $T_R$.
The asymmetry $Y_L$ at a certain temperature $T$ is 
estimated by solving the Boltzmann equations
\begin{eqnarray}
&&\frac{dY_{N_{1}}}{dz}=-\frac{z}{sH(M_1)}
\left(\frac{Y_{N_{1}}}{Y_{N_{1}}^{\rm eq}}-1\right)\left\{
\gamma_D^{N_1}+\sum_{j=1,2}\left(\gamma_{{N_1}{N_j}}^{(2)}+
\gamma_{{N_1}{N_i}}^{(3)}\right)\right\}, \nonumber \\
&&\frac{dY_L}{dz}=\frac{z}{sH(M_1)}\left\{
\varepsilon\left(\frac{Y_{N_{1}}}{Y_{N_{1}}^{\rm eq}}-1\right)\gamma_D^{N_1}
-\frac{2Y_L}{Y_\ell^{\rm eq}}\left(\gamma_N^{(2)} +\gamma_N^{(13)}\right)\right\}, 
\label{bqn}
\end{eqnarray}
where $z$ is defined as $z=\frac{M_1}{T}$.
The relevant reaction density $\gamma$ used in these equations 
can be found in \cite{ks}.\footnote{Since the lepton number violating 
effect due to sphaleron is not contained in eq.~(\ref{bqn}),
$Y_L$ in eq.~(\ref{bqn}) should be understood as $-Y_{B-L}$.
Non-degenerate right-handed neutrinos assumed here make 
the inverse decay to $N_{2,3}$ irrelevant in this analysis.}
The baryon number asymmetry in the present Universe is converted 
from this lepton number asymmetry $Y_L$ by the sphaleron interaction.
It can be estimated as $Y_B=-\frac{8}{23}Y_L(z_{\rm EW})$,
where the sphaleron decoupling temperature $T_{\rm EW}$ is taken to 
be 100~GeV. 

\begin{figure}[t]
\begin{center}
\epsfxsize=7.5cm
\leavevmode
\epsfbox{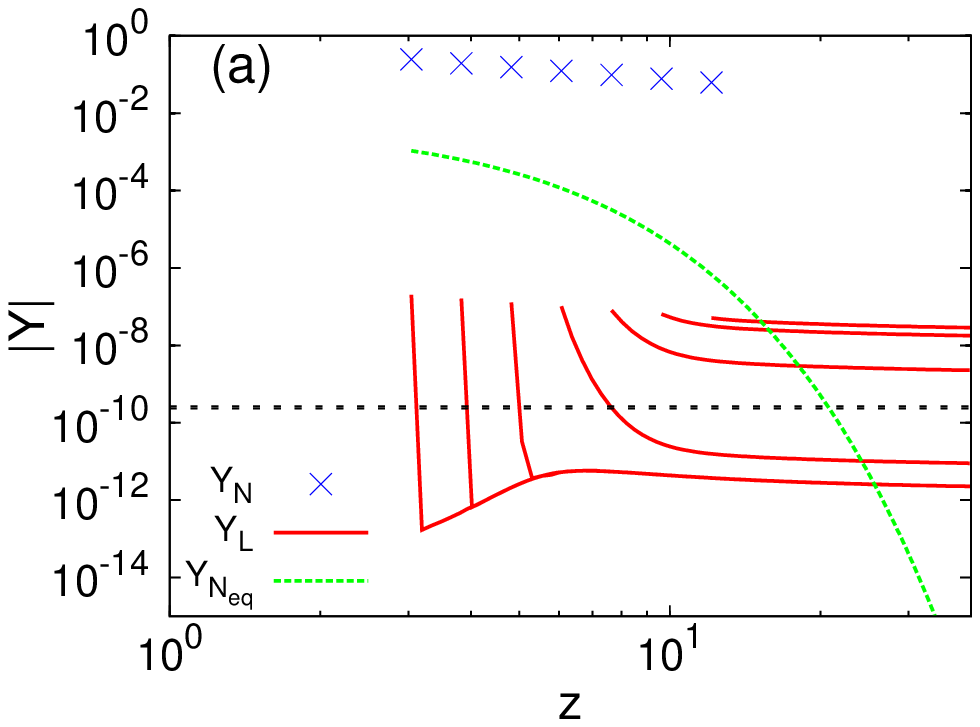}
\hspace*{5mm}
\epsfxsize=7.5cm
\leavevmode
\epsfbox{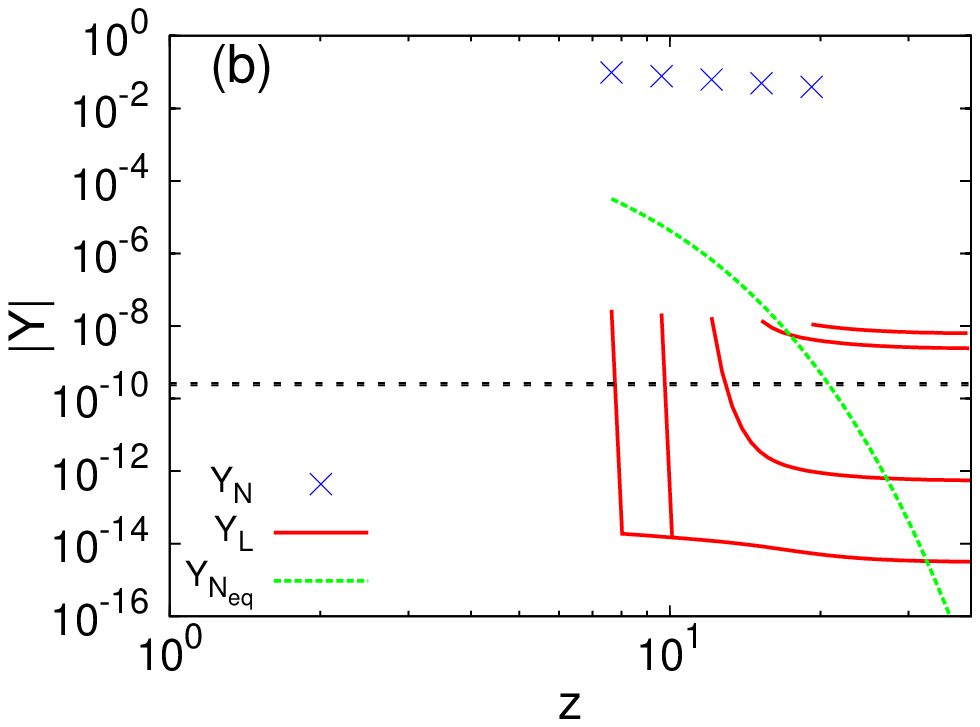}
\end{center}
\vspace*{-1mm}

{\footnotesize {\bf Fig.~2}~~The lepton number asymmetry $Y_L$ generated for
the parameter sets (a) and (b) in the neutrino sector shown in Table~1.
Crosses represent the initial value of $Y_{N_1}$ generated from the
inflaton decay. Horizontal dotted lines correspond to the lepton 
asymmetry required to explain the amount of the baryon number in the Universe.
Both $y_1$ and $m_{S_{\rm inf}}$ are taken so as to satisfy the 
conditions (\ref{inst}) and
(\ref{supp}). For each lines and crosses for $Y_L$ and $Y_{N_1}$, 
the value of $y_1$ is taken 
from left to right as $10^{-7.2},~ 10^{-7.3},
10^{-7.4},~ 10^{-7.5},~10^{-7.6},~10^{-7.7},~10^{-7.8}$ for $m_{S_{\rm inf}}=10^9$~GeV 
in (a), and $10^{-8.6},~10^{-8.7}, 10^{-8.8},~10^{-8.9},~10^{-9}$ 
for $m_{S_{\rm inf}}=10^7$~GeV in (b).}
\end{figure}
 
The model parameters used in the numerical analysis are given in Table~1.
They can explain well the neutrino oscillation data except for the
nonzero mixing angle $\theta_{13}$ as addressed before. 
In this study, we fix the initial 
values of $Y_{N_1}$ and $Y_L$ to be the thermal one at $T_R$ and 
the value fixed by eq.~(\ref{ini-yl}), respectively. 
Typical numerical results of the analysis are shown in Fig.~2.
Although $z_{\rm EW}$ is much larger than 20 in the assumed value of $M_1$,
$Y_L$ converges to a constant value at $z=20$ sufficiently, 
and then we can identify $Y_L(20)$ with the one at $z_{\rm EW}$.
The steep decrease of $Y_L$ at the initial stage of the evolution for a
larger $y_1$ is considered to be caused by the washout. 
It is effective for the larger $y_1$, which results in the higher 
reheating temperature as $T_R~{^>_\sim}~ M_1$. 
As the temperature decreases from $T_R$, the washout process is suppressed 
and $Y_L$ converges to a constant value as in case of the ordinary 
thermal leptogenesis.
For the smaller $y_1$, $T_R$ becomes sufficiently lower and 
the washout process is frozen. In that case, the required $Y_L$ can 
be obtained as long as its initial value is large enough.   
This result confirms that the reheating temperature and the decoupling 
of the washout effect are essential for the present scenario.

Since the initial value of $Y_L$ is determined by the neutrino Yukawa 
couplings $h_{1,2}$ which are constrained by the neutrino oscillation data, 
the scenario is closely related to the neutrino mass generation 
as in the ordinary leptogenesis.
However, we should also note that the model has an additional 
parameter $\lambda_5$ related to the neutrino mass. 
It makes the weak scale leptogenesis feasible also.  
If $|\lambda_5|$ takes a smaller value for fixed values of $M_i$,
the neutrino oscillation data require the larger neutrino Yukawa couplings 
$|h_i|$. In that case, the initial lepton number asymmetry becomes larger
but the washout effects become also stronger.
This suggests that a favorable value of $\lambda_5$ might be determined 
from a viewpoint of leptogenesis. 
Since $\lambda_5$ is also related to the DM physics in this model \cite{ks}, 
further study in this direction may give us a useful hint for the model. 

\section{Summary}  
We have proposed a scenario for the generation of the baryon number asymmetry
in a one-loop radiative neutrino mass model extended by 
the singlet scalars.
In this model, singlet scalars are related to both the inflation and 
the neutrino mass generation. 
Leptogenesis is caused by the decay of non-thermal right-handed 
neutrinos which is produced through the decay of inflaton.
If the right-handed neutrinos could decay immediately before they are 
thermalized, the lepton number asymmetry could be generated effectively
through this decay.
The number density of the non-thermal right-handed neutrino 
could be much larger than the thermal one so that the generated lepton 
asymmetry could be enhanced compared with the one which is generated 
from the decay of the thermal right-handed neutrinos.
Based on this lepton asymmetry, sphaleron could generate a 
sufficient amount of the baryon number asymmetry. 
We discussed the condition for which the non-thermal right-handed 
neutrinos could be the mother field of the lepton number asymmetry.
Numerical analysis for the evolution of the lepton number asymmetry 
shows that the sufficient baryon number asymmetry can 
be obtained from the decay of the right-handed neutrino, 
which is much lighter than the Davidson-Ibarra bound. 
Rather low reheating temperature could be sufficient for the 
generation of the required amount of the baryon number asymmetry 
in this scenario. 

\vspace*{5mm}
\section*{Acknowledgements}
This work is supported by MEXT Grant-in-Aid 
for Scientific Research on Innovative Areas (Grant Number 26104009).

\newpage
\bibliographystyle{unsrt}

\end{document}